\documentstyle[11pt,twoside]{article}

\begin{document}

\newcommand{\dd}{\,{\rm d}}
\newcommand{\ie}{{\it i.e.},\,}
\newcommand{\etal}{{\it et al.\ }}
\newcommand{\eg}{{\it e.g.},\,}
\newcommand{\cf}{{\it cf.\ }}
\newcommand{\vs}{{\it vs.\ }}
\newcommand{\zdot}{\makebox[0pt][l]{.}}
\newcommand{\up}[1]{\ifmmode^{\rm #1}\else$^{\rm #1}$\fi}
\newcommand{\dn}[1]{\ifmmode_{\rm #1}\else$_{\rm #1}$\fi}
\newcommand{\upd}{\up{d}}
\newcommand{\uph}{\up{h}}
\newcommand{\upm}{\up{m}}
\newcommand{\ups}{\up{s}}
\newcommand{\arcd}{\ifmmode^{\circ}\else$^{\circ}$\fi}
\newcommand{\arcm}{\ifmmode{'}\else$'$\fi}
\newcommand{\arcs}{\ifmmode{''}\else$''$\fi}
\newcommand{\MS}{{\rm M}\ifmmode_{\odot}\else$_{\odot}$\fi}
\newcommand{\RS}{{\rm R}\ifmmode_{\odot}\else$_{\odot}$\fi}
\newcommand{\LS}{{\rm L}\ifmmode_{\odot}\else$_{\odot}$\fi}

\newcommand{\Abstract}[2]{{\footnotesize\begin{center}ABSTRACT\end{center}
\vspace{1mm}\par#1\par
\noindent
{\bf Key words:~~}{\it #2}}}

\newcommand{\TabCap}[2]{\begin{center}\parbox[t]{#1}{\begin{center}
  \small {\spaceskip 2pt plus 1pt minus 1pt T a b l e}
  \refstepcounter{table}\thetable \\[2mm]
  \footnotesize #2 \end{center}}\end{center}}

\newcommand{\TableSep}[2]{\begin{table}[p]\vspace{#1}
\TabCap{#2}\end{table}}

\newcommand{\TableFont}{\footnotesize}
\newcommand{\TableFontIt}{\ttit}
\newcommand{\SetTableFont}[1]{\renewcommand{\TableFont}{#1}}

\newcommand{\MakeTable}[4]{\begin{table}[htb]\TabCap{#2}{#3}
  \begin{center} \TableFont \begin{tabular}{#1} #4 
  \end{tabular}\end{center}\end{table}}

\newcommand{\MakeTableSep}[4]{\begin{table}[p]\TabCap{#2}{#3}
  \begin{center} \TableFont \begin{tabular}{#1} #4 
  \end{tabular}\end{center}\end{table}}

\newenvironment{references}%
{
\footnotesize \frenchspacing
\renewcommand{\thesection}{}
\renewcommand{\in}{{\rm in }}
\renewcommand{\AA}{Astron.\ Astrophys.}
\newcommand{\AAS}{Astron.~Astrophys.~Suppl.~Ser.}
\newcommand{\ApJ}{Astrophys.\ J.}
\newcommand{\ApJS}{Astrophys.\ J.~Suppl.~Ser.}
\newcommand{\ApJL}{Astrophys.\ J.~Letters}
\newcommand{\AJ}{Astron.\ J.}
\newcommand{\IBVS}{IBVS}
\newcommand{\PASP}{P.A.S.P.}
\newcommand{\Acta}{Acta Astron.}
\newcommand{\MNRAS}{MNRAS}
\renewcommand{\and}{{\rm and }}
\section{{\rm REFERENCES}}
\sloppy \hyphenpenalty10000
\begin{list}{}{\leftmargin1cm\listparindent-1cm
\itemindent\listparindent\parsep0pt\itemsep0pt}}%
{\end{list}\vspace{2mm}}

\def\TYLDA{~}
\newlength{\DW}
\settowidth{\DW}{0}
\newcommand{\dw}{\hspace{\DW}}

\newcommand{\refitem}[5]{\item[]{#1} #2%
\def\REFARG{#3}\ifx\REFARG\TYLDA\else, {\it#3}\fi
\def\REFARG{#4}\ifx\REFARG\TYLDA\else, {\bf#4}\fi
\def\REFARG{#5}\ifx\REFARG\TYLDA\else, {#5}\fi.}

\newcommand{\Section}[1]{\section{#1}}
\newcommand{\Subsection}[1]{\subsection{#1}}
\newcommand{\Acknow}[1]{\par\vspace{5mm}{\bf Acknowledgments.} #1}
\pagestyle{myheadings}


\def\thefootnote{\fnsymbol{footnote}}

\begin{center}
{\Large\bf Metallicity of Red Clump Giants\\}
\vskip3pt
{\Large\bf in Baade's Window}
\vskip1cm
{\bf B~o~h~d~a~n~~P~a~c~z~y~{\'n}~s~k~i}
\vskip6mm
{Princeton University Observatory, Princeton, NJ 08544-1001, USA\\
e-mail: bp@astro.princeton.edu}
\end{center}
\vskip1cm
\Abstract{
The red clump giants are potentially very useful as standard candles.
There is some controversy about the stability of their I-band
absolute magnitude, but it does not seem to be serious.
No controversy was anticipated about their
colors, with metal rich giants expected to be redder and cooler than the
metal poor giants.  The purpose of this paper is to point out
that no such correlation is apparent between [Fe/H] and $ {\rm T_{eff} } $
as determined with Washington CCD photometry for the giants in Baade's
Window.  No explanation is offered for this surprising result.
It is also unknown why the galactic bulge red clump giants are redder
than the clump giants near the Sun by 0.2 mag in the $ {\rm (V-I)_0 } $ color.
} {Galaxy: center -- Hertzsprung-Russel (HR) diagram -- Stars: abundances --
Stars: fundamental properties -- Techniques: photometric}

\Section{Introduction}

Following the publication of the Hipparcos catalogue of parallaxes 
(Perryman \etal 1997) it became apparent that the red clump giants
near the Sun occupy a very small region in the color -- absolute magnitude
diagram, and therefore they are potentially very good standard candles.
Paczy\'nski and Stanek (1998) pointed out that red clump giants measured by
OGLE in Baade's Window (Udalski \etal 1993, Kiraga, Paczy\'nski and
Stanek 1997), and corrected for the interstellar reddening have remarkably
constant I--band magnitude over a large range of colors, 
$ {\rm 0.8 \leq (V-I)_0 \leq 1.4 } $.  Assuming that the absolute I-band
magnitude of red clump stars is the same near the Sun and in the
galactic bulge Paczy\'nski and Stanek (1998) determined the distance
to the galactic center to be $ {\rm R_0 = 8.4 \pm 0.4 ~ kpc } $, later
corrected to $ {\rm R_0 = 8.2 \pm 0.2 ~ kpc } $ by Stanek and Garnavich (1998).

Subsequently, the red clump stars were used to measure distance to M31 
(Stanek and Garnavich 1998), to both Magellanic Clouds (Udalski \etal 1998),
and to the Large Magellanic Cloud (Stanek, Zaritsky and Harris 1998). 
Udalski (1998a) demonstrated that distances based on red clump stars
and on RR Lyrae stars are compatible with each other, and that 
$ {\rm M_I } $ for the red clump stars depends weakly on their metallicity.
Finally, Udalski (1998b) showed that the absolute I-band magnitude of red
clump stars in the LMC and SMC star clusters is independent of age in the 
range $ 2 - 10 $ Gyr.  The last two papers presented empirical evidence
that red clump stars can be used as standard candles to determine distance 
modulus with an accuracy of 0.1 mag, or perhaps even better.  However,
care must be taken to estimate the age and metallicity of the population 
used for the study.
Note, that field red clump giants in both Clouds have the same
I-band magnitudes and colors as the red clump giants in star clusters 
which are 2 -- 10 Gyr old (Udalski 1998a,b, Udalski \etal 1998).

Theoretical models indicate that red clump stars have a small range
of absolute luminosities (Seidel, Demarque, and Weinberg 1989, Castellani,
Chieffi, and Straniero 1992, Jimenez, Flynn, and Kotoneva 1997).
The question is: how small?
Recently Cole (1998) and Girardi \etal (1998) pointed out that theoretically
expected range of I-band luminosities may be as large as 0.5 mag,
and that the systematic difference between the corresponding stellar
populations makes the Magellanic Clouds more distant than claimed
by Udalski \etal (1998).  The apparent conflict is not likely to be 
serious, when various theoretical uncertainties are
taken into account.  In particular the age distribution
and helium abundance adopted in theoretical models may be inaccurate.

The aim of this paper is to point out that while there is a debate about
the stability of the $ {\rm M_I } $ magnitude of the red clump stars, there is
a another serious problem with the interpretation of their $ {\rm (V-I)_0 } $
colors, and very likely with the [Fe/H] values as determined with Washington
CCD photometry.  No explanation for the apparent problem is offered in this
paper.

\Section{Theoretical expectations}

It is well established that when metallicity is increased the opacity
increases too, and in particular the opacity in a K giant atmosphere
goes up.  As such giants have deep convective envelopes the increase
of atmospheric opacity makes stellar radii larger, and their effective
temperatures lower.  This theoretical result was stable over last several
decades, and it is well presented in Fig. 2a of Jimenez, Flynn and
Kotoneva (1997).  As far as I know the notion that the more metal rich
giants are cooler is very robust and well understood, and it has never
been challenged.

The effective temperature has a decisive effect on stellar colors.
Higher metallicity implies redder color for two reasons.  First, because
the higher the metallicity, the lower the temperature, and the redder the 
color.  Second, because the B-band (and even more so the U-band) are 
affected by line blanketing, and the blanketing makes colors redder.
However, as both effects work in the same direction the prediction
is clear: the more metal rich stars should be redder (Jimenez \etal
1997, Girardi \etal 1998, and references therein).

The age of a population may affect the colors as well, but the effect
is modest, and it is very small for K giants older than 2 Gyr
(Jimenez \etal 1997, Fig. 3, Girardi \etal 1998, Fig. 1).

The $ {\rm (V-I)_0 } $ color is almost unaffected by line blanketing,
so it should be a good temperature indicator.  The Washington
index $ {\rm (T_1-T_2 ) } $ is claimed to be even better temperature
indicator (Taylor \etal 1987).  Therefore, we expect a good correlation
between K-giant metallicity and the suitable color index.  

\Section{Galactic bulge}

This expectation that $ {\rm (V-I)_0 } $ color is well correlated with
metallicity was the basis for the suggestion by Paczy\'nski and Stanek
(1998) that the galactic bulge red clump stars are more metal rich
than their counterparts near the Sun, as the average colors are
$ {\rm \langle (V-I)_0 \rangle \approx 1.2 } $ in the Baade's Window bulge
and $ {\rm \langle (V-I)_0 \rangle \approx 1.0 } $ near the Sun.

\begin{figure}[t]
\vspace{8.0cm}
\includegraphics{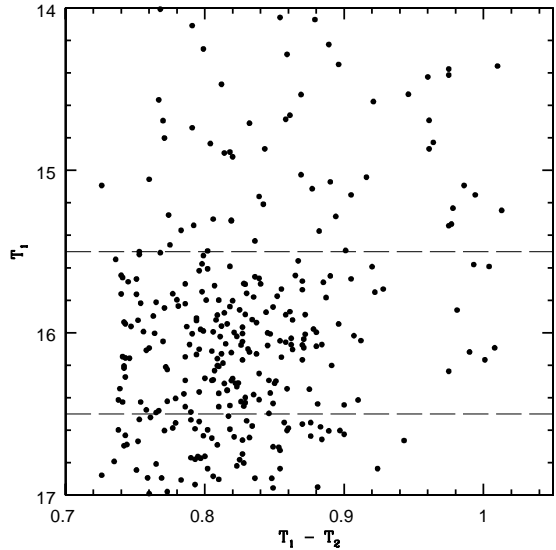}
\caption{
Color -- magnitude diagram for K giants in Baade's Window based on
Washington CCD photometry presented in Table 1 of Geisler and Friel (1992).
The region between the two horizontal dashed lines corresponds to the
red clump giants.
}
\end{figure}

The metallicity of the galactic bulge giants has
a long and complicated history.  For many years the bulge giants were thought
to be super metal rich (cf. Geisler and Friel 1992, and references 
therein), but the estimate of their [Fe/H] was recently reduced 
(cf. Sadler, Rich, and Terndrup 1996, and references therein).
The range of bulge metallicities is still believed to be very large.
Note, that the range of $ {\rm (V-I)_0 } $
colors for the bulge red clump giants is about twice larger than the
corresponding range near the Sun.

It seems that the mean metallicity is currently thought to be about the same 
in the bulge and near the Sun.  Yet, the bulge red clump giants are on
average redder by 0.2 mag than the red clump giants near the Sun.  These
two statements appear to be in conflict with the theoretical expectations
presented in the previous chapter. 

\begin{figure}[p]
\vspace{8.0cm}
\includegraphics{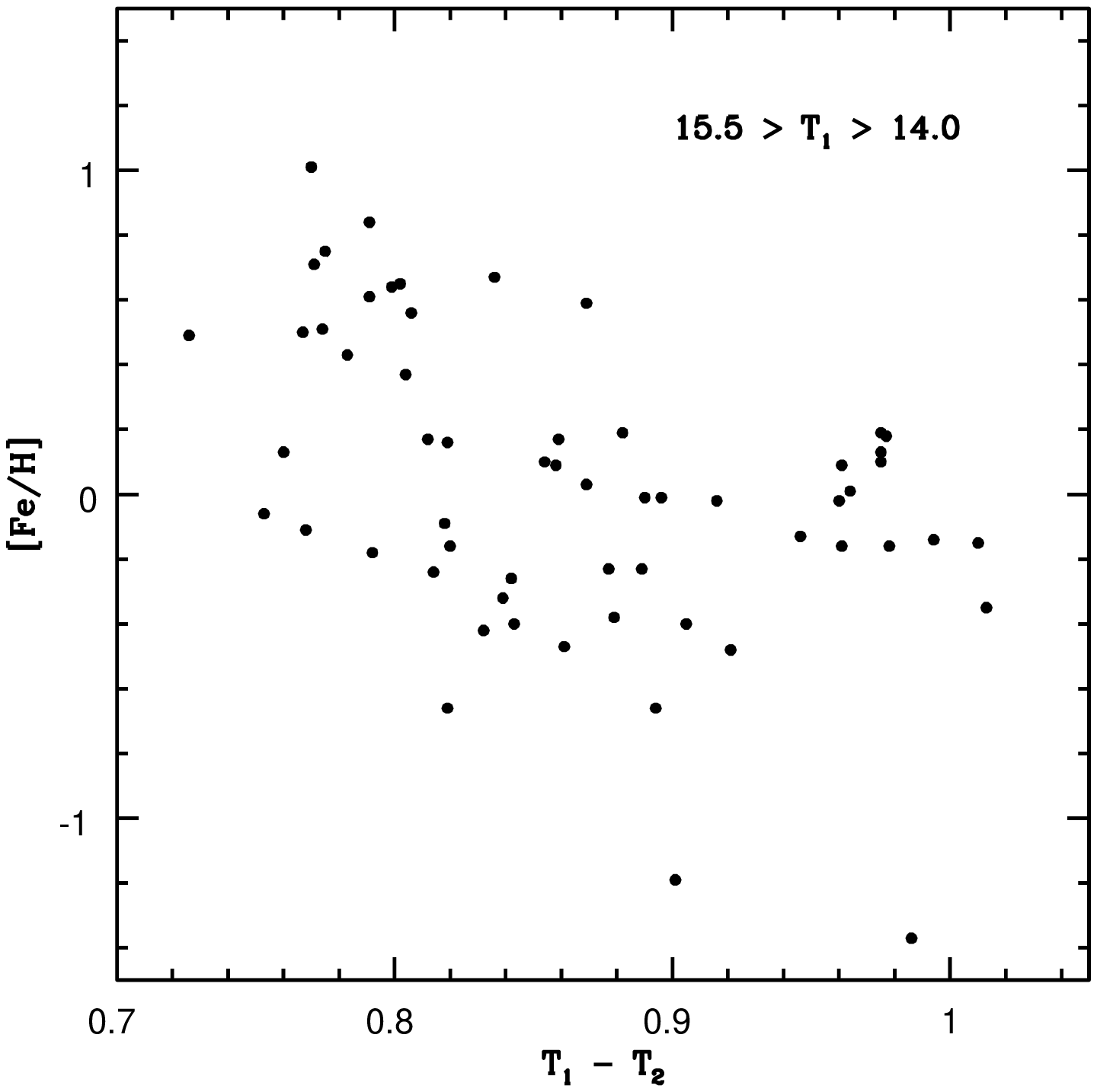}
\caption{
The [Fe/H] -- color diagram diagram for K in Baade's Window based on
Washington CCD photometry presented in Table 1 of Geisler and Friel (1992).
The stars shown are brighter than red clump giants.  The apparent
correlation is opposite to the one expected, with the metal richer stars
being bluer.
}
\vspace{8.0cm}
\includegraphics{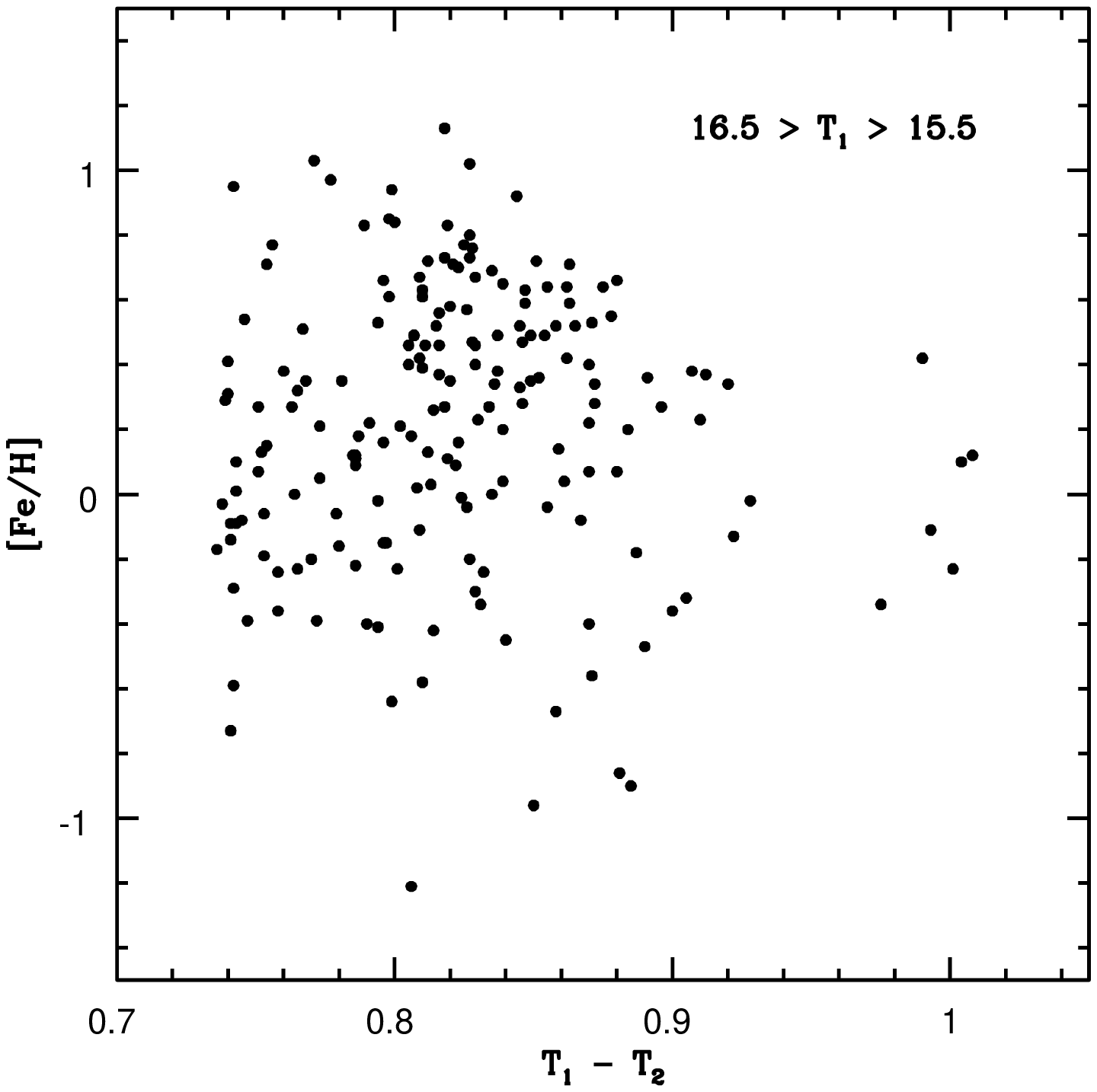}
\caption{
The [Fe/H] -- color diagram diagram for K giants in Baade's Window based on
Washington CCD photometry presented in Table 1 of Geisler and Friel (1992).
The stars shown are the red clump giants.  There is no apparent correlation 
between their metallicity and color.
}
\end{figure}

\begin{figure}[p]
\vspace{8.0cm}
\includegraphics{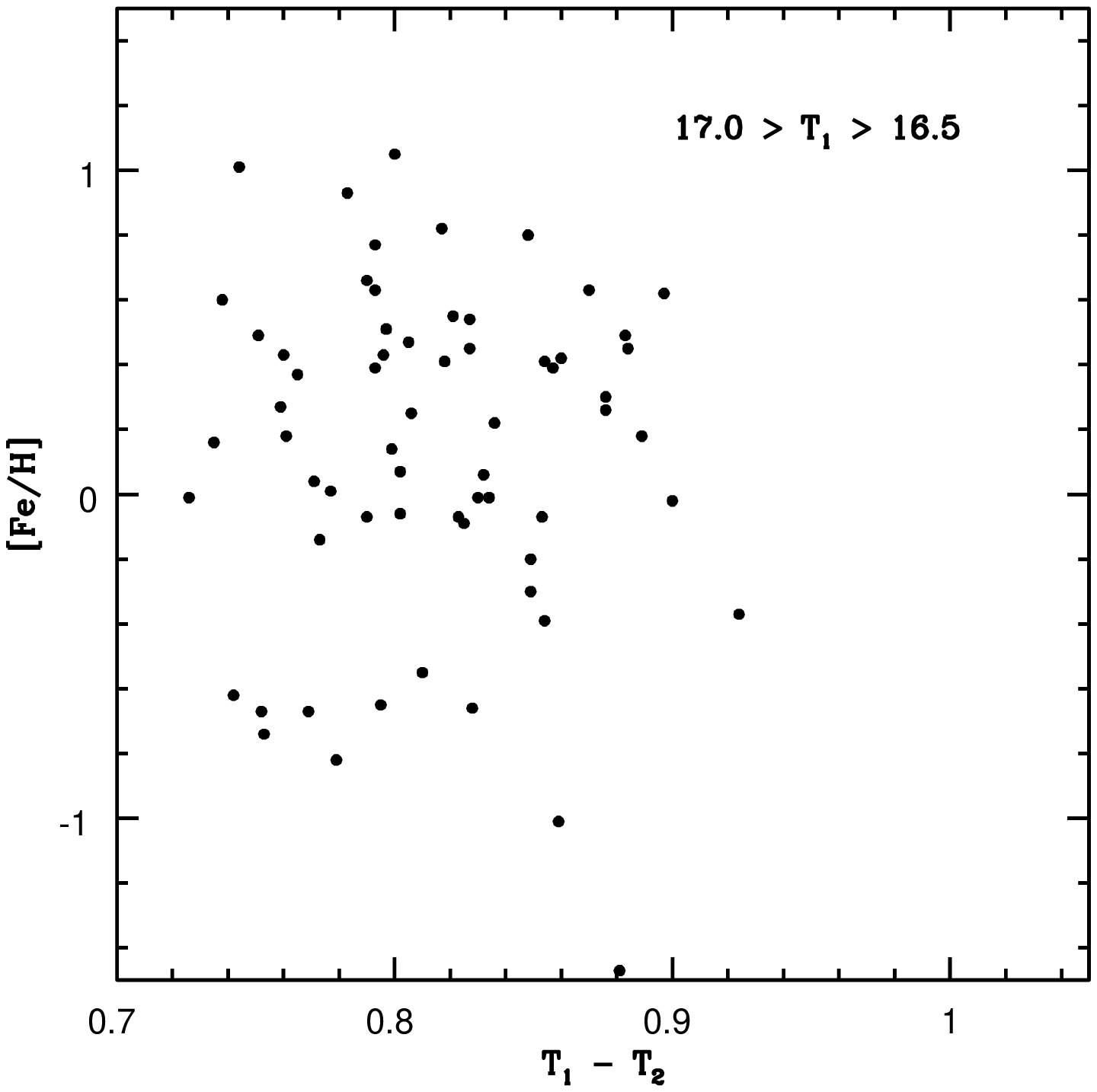}
\caption{
The [Fe/H] -- color diagram diagram for K giants in Baade's Window based on
Washington CCD photometry presented in Table 1 of Geisler and Friel (1992).
The stars shown are fainter than red clump giants.  There is no apparent
correlation between their metallicity and color.
}
\vspace{8.0cm}
\includegraphics{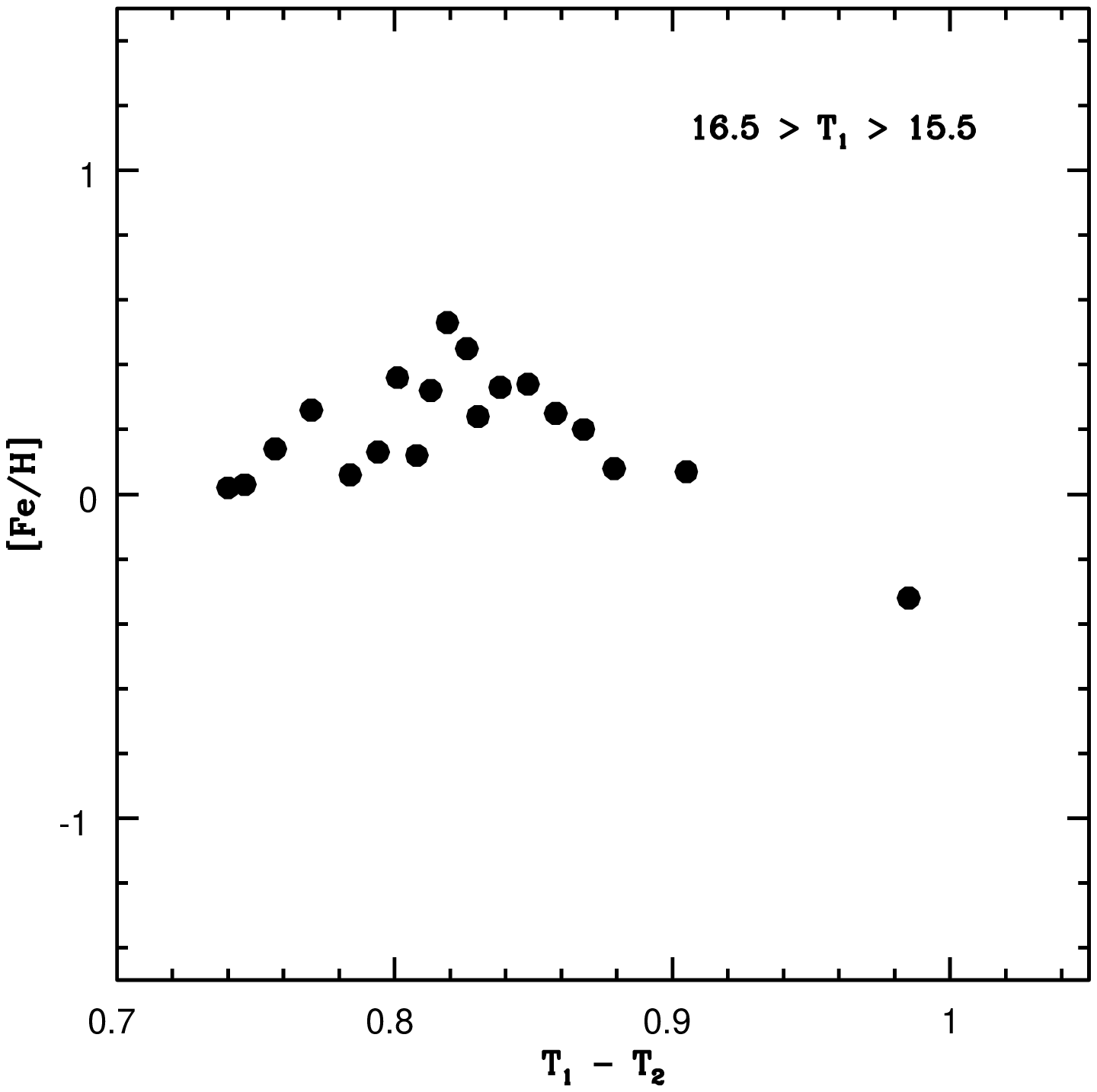}
\caption{
The [Fe/H] -- color diagram diagram for K giants in Baade's Window based on
Washington CCD photometry presented in Table 1 of Geisler and Friel (1992).
Each point corresponds to the average of 10 points in Fig. 3. 
No correlation between the metallicity and colors of the
red clump giants is apparent.
}
\end{figure}

Another check of the theoretical expectations is offered by the data
published by Geisler and Friel (1992, Table 1), who obtained [Fe/H]
and the $ {\rm (T_1-T_2) } $ temperature index for over 300 stars in
Baade's Window.  Their data is presented in a color - magnitude diagram 
in Fig. 1, which is a small part of Fig. 3 of Geisler and Friel (1992).
The red clump giants are located between the two horizontal dashed lines.
The area in the sky covered was only $ (3')^2 $, so the interstellar
extinction was more or less uniform, and the $ {\rm (T_1-T_2) } $ index
should be well correlated with the effective temperature.  All data
from Table 1 of Geisler and Friel (1992) are plotted in Figs. 2-4 in
$ {\rm (T_1-T_2) - [Fe/H] } $ diagrams, for the stars above the red clump
(Fig. 2), in the clump (Fig. 3), and below the clump (Fig. 4).  A strong
positive correlation was expected theoretically in all three figures.
The only correlation which is clearly apparent is {\it negative} for
giants above the red clump: in the magnitude range $ 15.5 > T_1 > 14.0 $
the metal richer giants are bluer!  The red clump giants were binned
in groups of 10, and the binned [Fe/H] values are plotted versus temperature
index in Fig. 5.   Again, no correlation is apparent.

\Section{Discussion}

Something is very wrong either with our understanding of stellar structure,
or with the determination of [Fe/H] with Washington CCD photometry,
or with the color index $ {\it (T_1-T_2) } $ as temperature indicator.

Recently, H${\rm \o }$g and Flynn (1997) presented DDO photometry for Hipparcos
K giants with [Fe/H] $ > - 0.5 $.  Dr. Flynn kindly provided me with the 
$ {\rm (B-V) - [Fe/H] } $ and $ {\rm (V-I) - [Fe/H] } $ diagrams for the
Hipparcos red clump giants.  There was a clear correlation apparent on the 
first diagram, but none was obvious in the second.  It is possible that
line blanketing which is strong in the B-band was at least partly
responsible for the correlation between $ {\rm (B-V) } $ and [Fe/H].
It is surprising that the correlation between $ {\rm (V-I) } $ and [Fe/H]
is either very weak or absent.

Perhaps this indicates that [Fe/H] is not a good measure of the true
stellar metallicity?  Perhaps, but this is not the whole story.
An inspection of Figures 2a and 3 of Jimenez \etal (1997) and Fig. 1 of 
Girardi \etal (1998) makes it clear that according to theoretical
models the metallicity, not age or stellar mass is decisive in
establishing the colors and the range of colors of the red clump giants.
In particular, the only way to obtain $ {\rm (V-I)_0 \approx 1.4 } $,
as observed in Baade's Window, is by adopting $ {\rm Z \gg 0.03 } $, 
The observed fact that the bulge red clump giants are
redder then the red clump giants near the Sun by 0.2 mag, on average,
requires their metallicity to be higher by a factor $ \sim 4 $.  This
is in a direct conflict with the recent spectroscopic determinations
(cf. Sadler, Rich, and Terndrup, 1996, and references therein).

According to Udalski (1998a,b) and Udalski et al. (1998) the red clump giants
in the Magellanic Clouds and in the Carina dwarf galaxy have colors
$ {\rm \langle (V-I)_0 \rangle _{LMC} \approx 0.90 } $,
$ {\rm \langle (V-I)_0 \rangle _{SMC} \approx 0.85 } $,
$ {\rm \langle (V-I)_0 \rangle _{Car} \approx 0.82 } $,
while their [Fe/H] is estimated to be $ -0.6, ~ -0.8, ~ -1.9 $,
respectively.  There
is a trend between the color and the metallicity, which is monotonic
all the way to the red clump giants near the Sun.  Only
the galactic bulge stars appear to suffer the lack of correlation
between their metallicity and colors.

It is possible that there is no single parameter which can be used
to describe `metallicity', perhaps the elemental ratios vary a lot.
Perhaps some elements are responsible for continuum atmospheric
opacity, and therefore broad band colors, and some other are 
responsible for the line features used to measure `metallicity'
(Paczy\'nski and Stanek, 1998)?

Following a lengthy discussion of this dilemma with a number of people
listed in the acknowledgments I was unable to convince myself that the
reason for the apparent conflict can be explained. Theoretically, a strong
correlation between the $ {\rm (V-I)_0 } $ and/or $ {\rm (T_1-T_2) } $ colors
and the [Fe/H] metallicity index is expected, but none is detectable for
the red clump giants either in the galactic bulge or near the Sun.

Note, that absolute I-band luminosity of the red clump stars
depends on their metallicity, as shown by the data (Udalski 1998a)
and the models (Jimenez \etal 1997, Girardi \etal 1998).  Therefore,
a clear and unambiguous method to determine their metallicity is
needed if they are to be used as reliable standard candles.

\Acknow{
It is a great pleasure to acknowledge numerous e-discussions with Dr. Dr.
C. Flynn, L. Girardi, R. Jimenez, D. Terndrup, A. Udalski, and A. Weissr,
and the hospitality by the personnel of the Institut d'Astrophysique, CNRS, 
where this paper was written.
This work was supported with the NSF grants AST--9313620 and AST--9530478.}


\end{document}